\documentclass[%
 reprint,
superscriptaddress,
 amsmath,amssymb,
 aps,
]{revtex4-1}

\usepackage{graphicx}
\usepackage{dcolumn}
\usepackage{bm}
\usepackage[usenames,dvipsnames]{color}
\usepackage{hyperref}
\usepackage{bbold}
\usepackage{lineno}

\begin{document}

\title{Experimental signature of a topological quantum dot}
\author{Marie S. Rider}
\email{marie.rider@ic.ac.uk}
\affiliation{The Blackett Laboratory, Imperial College London, London SW7 2AZ, United Kingdom}

\author{Maria Sokolikova}
\affiliation{Department of Materials, Imperial College London, London SW7 2AZ, United Kingdom}

\author{Stephen M. Hanham}
\affiliation{Department of Materials, Imperial College London, London SW7 2AZ, United Kingdom}
\affiliation{Department of Electronic, Electrical and Systems Engineering, University of Birmingham, Birmingham B15 2TT, United Kingdom}

\author{Miguel Navarro-C\'{i}a}
\affiliation{Department of Electronic, Electrical and Systems Engineering, University of Birmingham, Birmingham B15 2TT, United Kingdom}

\author{Peter Haynes}
\affiliation{Department of Materials, Imperial College London, London SW7 2AZ, United Kingdom}

\author{Derek Lee}
\affiliation{The Blackett Laboratory, Imperial College London, London SW7 2AZ, United Kingdom}

\author{Maddalena Daniele}
\affiliation{INFN and Department of Physics, Sapienza University of Rome, Piazzale A. Moro 2, 00185 Rome, Italy}

\author{Mariangela Cestelli Guidi}
\affiliation{INFN-LNF, Via Enrico Fermi 40, 00444 Frascati (Roma) Italy}

\author{Cecilia Mattevi}
\affiliation{Department of Materials, Imperial College London, London SW7 2AZ, United Kingdom}

\author{Stefano Lupi}
\affiliation{INFN and Department of Physics, Sapienza University of Rome, Piazzale A. Moro 2, 00185 Rome, Italy}

\author{Vincenzo Giannini}
\homepage{www.GianniniLab.com}
\affiliation{Instituto de Estructura de la Materia (IEM-CSIC), Consejo Superior de Investigaciones Cient{\'i}ficas, Serrano 121, 28006 Madrid, Spain}
\affiliation{The Blackett Laboratory, Imperial College London, London SW7 2AZ, United Kingdom}


\maketitle

\begin{figure}
\includegraphics[width=\columnwidth]{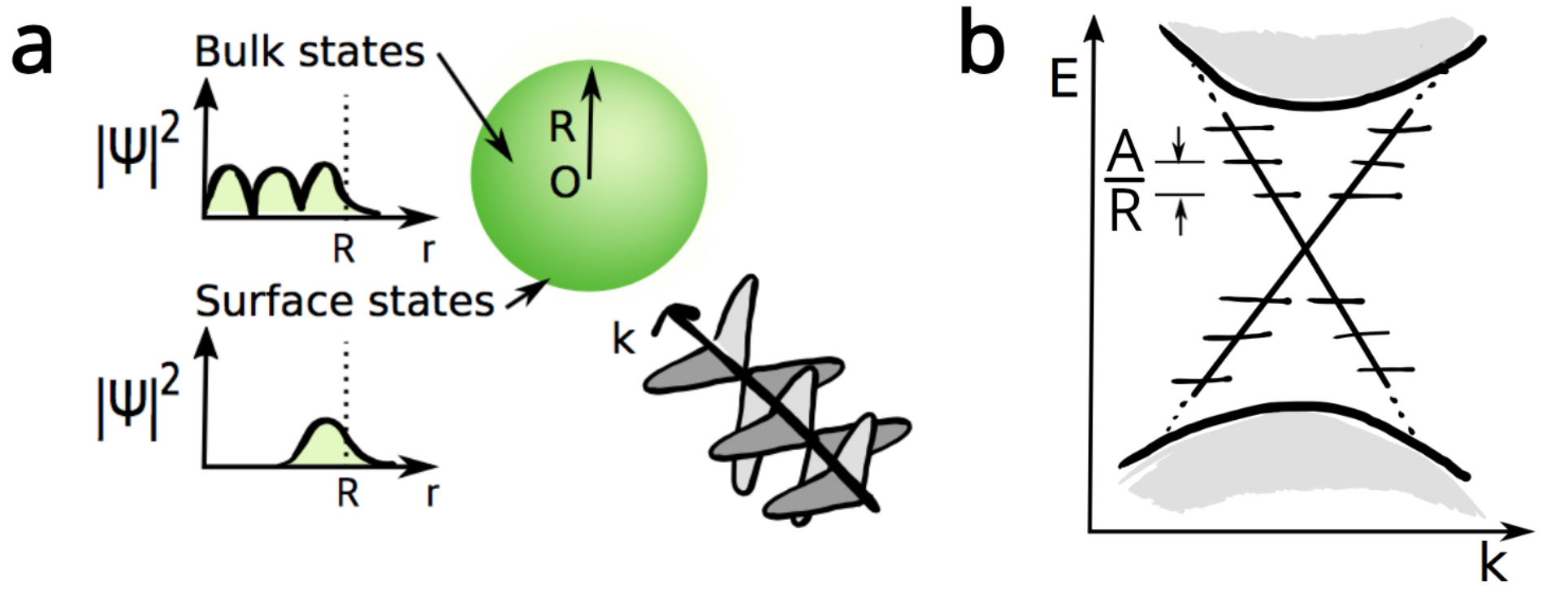}
\caption{\textbf{Small topological insulator nanoparticles}: \textbf{(a)} Schematic of TINP of radius $R$ irradiated with light, illustrating bulk states and surface states which decay into the bulk with a length scale $\sim$\AA.  \textbf{(b)} For small TINPs ($R<$100~nm) the Dirac cone becomes discretized with the position of energy levels inversely scaling with $R$.}\label{fig:schematic}
\end{figure}

\textbf{Topological insulators (TIs) present a neoteric class of materials, which support delocalised, conducting surface states despite an insulating bulk. Due to their intriguing electronic properties, their optical properties have received relatively less attention. Even less well studied is their behaviour in the nanoregime, with most studies thus far focusing on bulk samples - in part due to the technical challenges of synthesizing TI nanostructures. We study topological insulator nanoparticles (TINPs), for which quantum effects dominate the behaviour of the surface states and quantum confinement results in a discretized Dirac cone, whose energy levels can be tuned with the nanoparticle size. The presence of these discretized energy levels in turn leads to a new electron-mediated phonon-light coupling in the THz range. We present the experimental realisation of Bi$_2$Te$_3$ TINPs and strong evidence of this new quantum phenomenon, remarkably observed at room temperature. This system can be considered a topological quantum dot, with applications to room temperature THz quantum optics and quantum information technologies.}
\\
\\
Topological nanophotonics is an emerging discipline that aims to control light at the nanoscale using topological states~\cite{rider2019perspective}. Topological insulators are 3D materials which are insulating in the bulk but host topologically protected, conducting surface states~\cite{fu2007topological,chen2009experimental,moore2010birth,stauber2017plasmonics}. These surface states display a linear dispersion relation, seen as a Dirac cone which bridges the bulk band gap. The Dirac cone can only be opened by breaking time-reversal symmetry, making these states remarkably robust to perturbations such as material defects and stray (non-magnetic) fields. These surface states also do not exhibit back scattering. Pioneering work~\cite{imura2012spherical,zhang2012surface} has studied topological insulator particles, which are finite systems exhibiting the same level of topological protection as their continuous counterparts~\cite{PhysRevMaterials.1.024201}. In a bulk sample, only a small number of atoms in the material contribute to surface state behaviour and so topological effects are notoriously weak. By studying these materials in the nanoregime, there is an enhanced surface area to bulk ratio, and a larger proportion of atoms participate in the surface physics. This results in a pronounced topological contribution to the electronic and optical properties of the system. Some nanostructures (such as nanoribbons and thin films~\cite{ginley2018dirac}) can already be reliably produced, but spherical TINPs (whose surface to volume ratio is maximal whilst maintaining a true 3D `bulk') have so far proved challenging (a schematic of a TINP and its bulk and surface states is given in FIG. \ref{fig:schematic}a).
\\
%
%
As shown by Siroki \textit{et al.}~\cite{siroki2016single}, in the case of small ($5<R/\mathrm{nm}<100$ ) TINPs the continuous Dirac cone becomes discretized due to finite size effects, giving a set of discrete energy levels much like those in a quantum dot. The discretized energy levels of the surface states are equally spaced, with energies inversely proportional to $R$ (as illustrated schematically in FIG. \ref{fig:schematic}b). When irradiated with THz light, a transition between two of these topological, delocalised surface states occurs within the same frequency range as a bulk phonon excitation, resulting in a strong Fano resonance, referred to as the surface topological particle (SToP) mode~\cite{siroki2016single}. This is a purely quantum mechanical feature of the system, and the asymmetric profile of this resonance creates a point of zero-absorption when the energy spacing of the surface states is matched by the incident light, meaning that remarkably, the excitation of a single electron occupying a topological surface state can shield the bulk from the absorption of incoming light. This mode has been theoretically predicted~\cite{siroki2016single}, but had not been experimentally observed until the present work. Light-matter interactions in nanoparticles requiring a quantum mechanical approach have attracted much attention in quantum plasmonics in recent years~\cite{fitzgerald2016quantum}, and TINPs made of materials such as Bi$_2$Te$_3$ present a promising extension of this field.
\\ 

\begin{figure}[ht]
\includegraphics[width=\columnwidth]{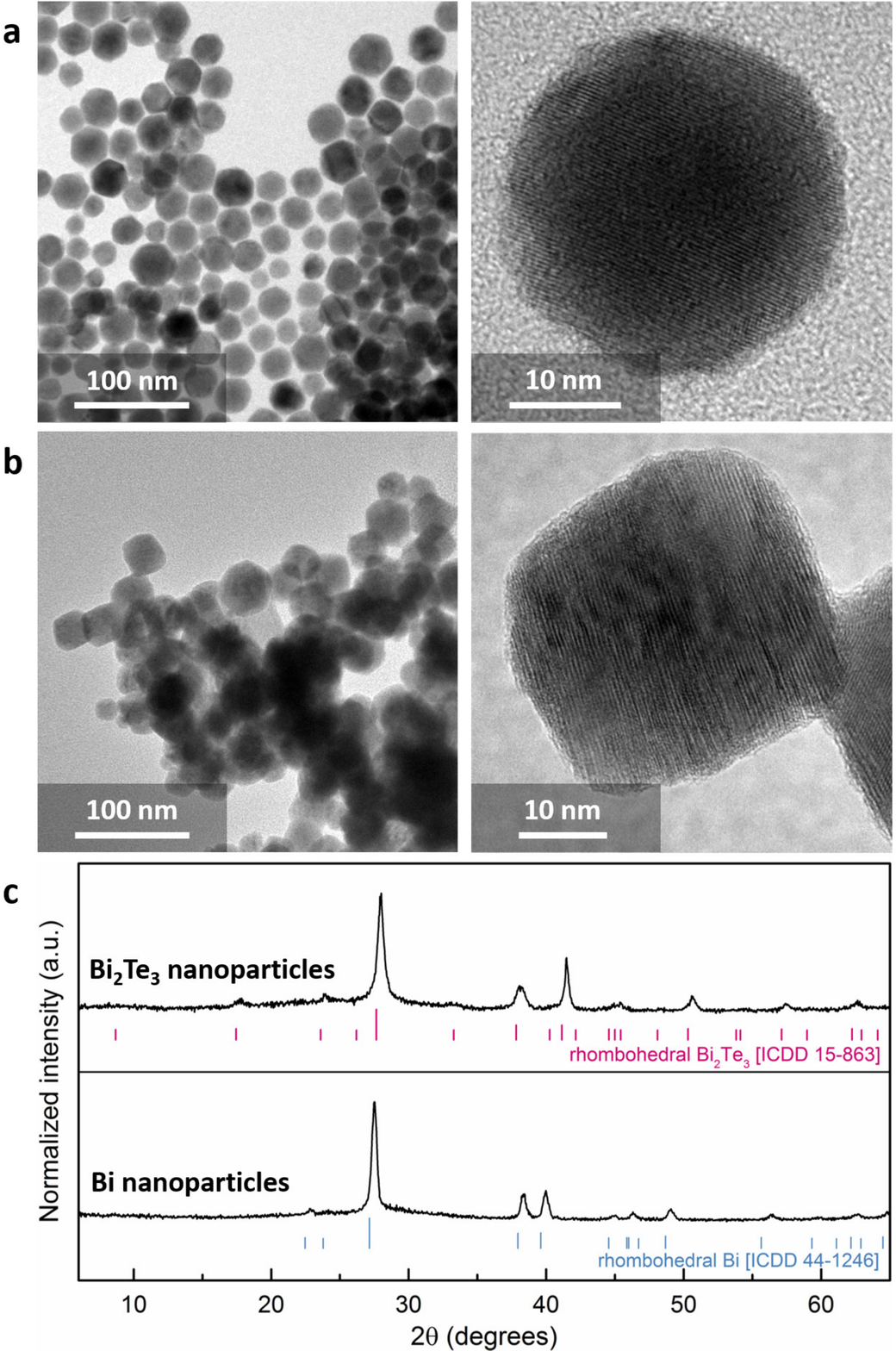}
\caption{\textbf{Bi$_2$Te$_3$ nanoparticle synthesis:} \textbf{(a)} TEM images of Bi nanoparticles, which are nearly spherical and of average radius 14.4 nm. \textbf{(b)} Successfully synthesized Bi$_2$Te$_3$ nanoparticles with a slightly rhombohedral shape and average radius of 17.5 nm. \textbf{(c)} X-ray diffraction patterns illustrating the successful tellurisation of the Bi nanoparticles to form Bi$_2$Te$_3$ nanoparticles.}\label{fig:method}
\end{figure}

Bi$_2$Te$_3$ has a layered crystal structure with quintuple covalently-bonded Te-Bi-Te-Bi-Te layers stacked along the {\bf c}-axis and therefore can be easily produced in the form of two-dimensional sheets via physical~\cite{teweldebrhan2010exfoliation} or wet-chemical  methods~\cite{sokolikova2017room}. In order to suppress preferential formation of anisotropic two-dimensional structures unavoidable in the direct solution-phase synthesis due to the high reactivity of Bi salts, spherical Bi$_2$Te$_3$ nanoparticles were synthesised using a two-stage reaction analogously to Ref~\cite{scheele2009synthesis}. Briefly, the growth involved the reduction of Bi ions by organic primary amine in an inert atmosphere leading to the formation of spherical Bi nanoparticles and followed by tellurisation of the produced Bi nanospheres in trioctylphosphine telluride solution. Transmission electron microscopy (TEM) images shown in FIGs. \ref{fig:method}a and \ref{fig:method}b are representative images of the ensembles of Bi and Bi$_2$Te$_3$ nanoparticles respectively. The Bi nanoparticles are nearly spherical with radius of $14.4\pm$2.3~nm, while upon tellurisation the shape of Bi$_2$Te$_3$ nanoparticles changes to slightly rhombohedral with a larger average radius of 17.5~nm (standard deviation 1.8~nm) due to the inclusion of tellurium species. Powder X-ray diffraction (XRD) patterns of both intermediate Bi and final Bi$_2$Te$_3$ nanoparticles clearly demonstrate a phase change accompanied by a significant shift of the diffraction peaks to the higher angles corresponding to smaller interplanar distances of the rhombohedral Bi$_2$Te$_3$ lattice (FIG.~\ref{fig:method}c). A more detailed experimental method is provided in Appendix \ref{app:exp}.
\\

%
\begin{figure*}
\includegraphics[width=\textwidth]{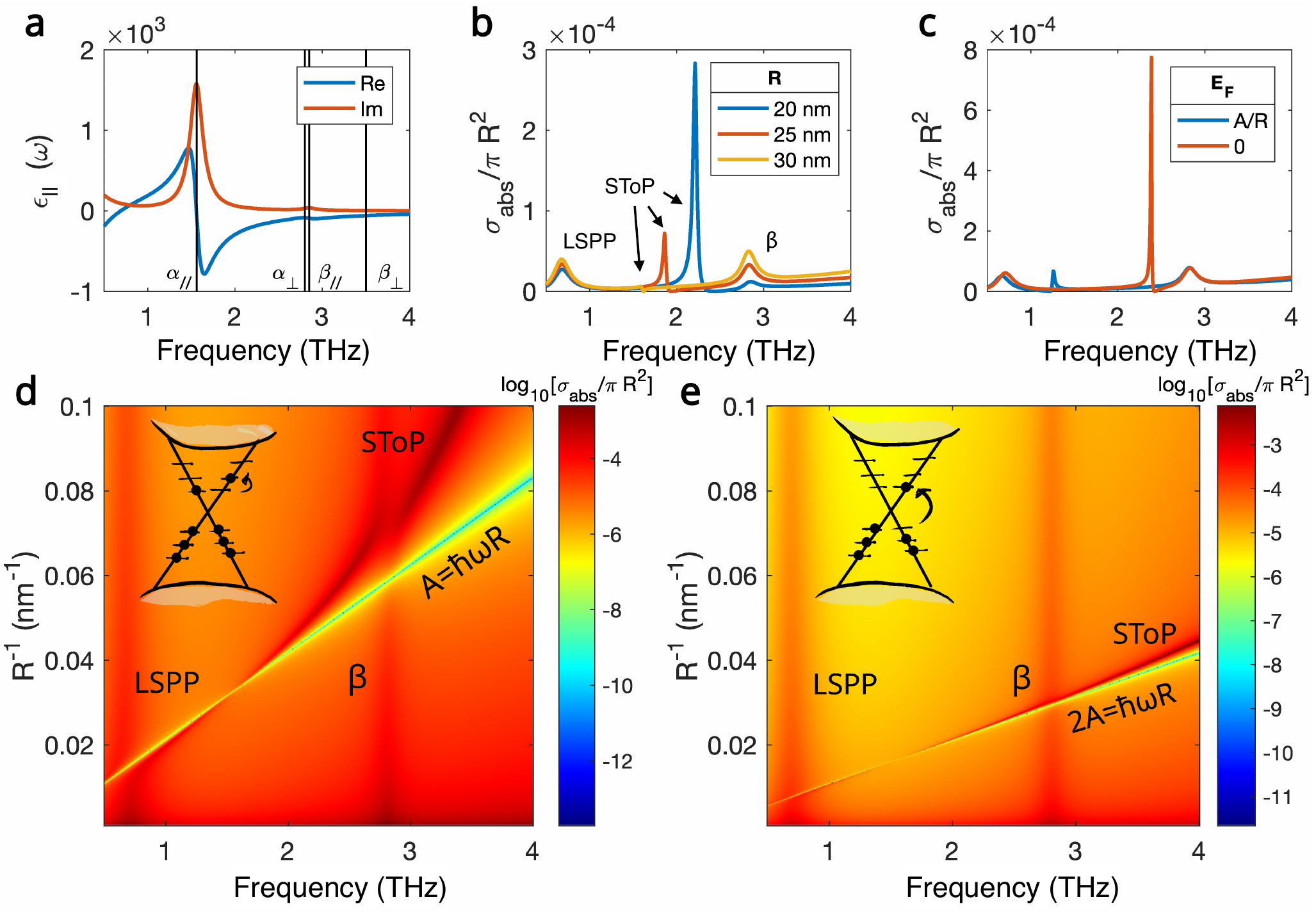}
\caption{\textbf{Theoretical SToP results:} \textbf{(a)} The real (blue) and imaginary (orange) components of the the bulk dielectric function $\epsilon_{\parallel}(\omega)$ describing the bulk optical response of Bi$_2$Te$_3$ irradiated with light propagating along its \textbf{c}-axis, using parameters fitted to data in ref~\cite{PhysRevB.96.235202}, presented in TABLE \ref{table:bi2te3}. The positions of the $\alpha$ and $\beta$ phonon frequencies are annotated for both \textbf{k}$\parallel$\textbf{c} and \textbf{k}$\perp$\textbf{c} incoming light (denoted $\alpha_{\parallel}$, $\beta{\parallel}$, $\alpha_{\perp}$, and $\beta{\perp}$ respectively). \textbf{(b)} Absorption cross-section of a TINP for three radii $R$=\{20, 25, 30\}~nm and Fermi level $E_\mathrm{F} = A/R$. The localised surface plasmon-polariton (LSPP) mode and the $\beta$ mode are visible and labelled, whilst the $\alpha$ mode is over-damped and not seen in the absorption cross-section. Increasing the TINP radius causes the SToP mode peak (also labelled) to occur at a lower frequency. For each SToP mode there is a point of zero absorption which occurs after the peak. \textbf{(c)} Changing the Fermi level from E$_\mathrm{F}$ = $A/R$ to E$_\mathrm{F}=0$ results in the SToP mode occurring at twice the frequency of incoming light, shown for $R=40$~nm. \textbf{(d)} Plot of the absorption cross-section for a range of $R$ and $\omega$, $E_\mathrm{F} = A/R$. The bulk LSPP and $\beta$ phonon modes are seen to remain constant for varying radius size, and  the line of zero-absorption is annotated ($A=\hbar \omega R$). \textbf{(e)} A similar plot for $E_{\mathrm{F}} = 0$. \label{fig:results}}
\end{figure*}
%
%
In the theoretical treatment of a Bi$_2$Te$_3$ TINP and incoming THz light of frequency $\omega/2\pi$, we treat the bulk behaviour classically with a bulk dielectric function and the surface states are treated quantum mechanically. We focus on Bi$_2$Te$_3$ but the following analytic results are valid for other materials within the Bi$_2$Te$_3$ family given the correct parameter substitutions. We work in the limit in which the particle radius $R$ is much smaller than the wavelength of incoming light $R \ll \lambda$. The bulk dielectric function of Bi$_2$Te$_3$ is a rank 2 tensor and to treat the system analytically we take $\epsilon (\omega)$ as a diagonal matrix with principle components $[\epsilon_{\perp}(\omega),\epsilon_{\perp}(\omega),\epsilon_{\parallel}(\omega)]$ and all other components equal to 0. $\epsilon_{\parallel}(\omega)$ gives the dielectric function along the \textbf{c}-axis of the material, while $\epsilon_{\perp}(\omega)$ is the dielectric function in both axes perpendicular to the \textbf{c}-axis. Each principle component is of the form
\begin{equation}
\epsilon(\omega) = \sum_{\mathrm{j}=\alpha,\beta,\mathrm{f}} \frac{\omega_{\mathrm{p,j}}^2}{\omega_{0,\mathrm{j}}^2-\omega^2-  2\pi i \gamma_\mathrm{j} \omega},
\label{eq:diElec}
\end{equation}  
which contains contributions from $\alpha$ and $\beta$ transverse phonons and free charge carriers (denoted by $\mathrm{f}$) arising from bulk defects. The parameters (given in TABLE \ref{table:bi2te3}) for $\epsilon_{\parallel}(\omega)$ have been determined by fitting to experimental data for samples illuminated with light propagating along the \textbf{c}-axis of the material~\cite{PhysRevB.96.235202}, and measured at 300~K. The real and imaginary parts of $\epsilon_{\parallel}(\omega)$ are plotted in FIG.~\ref{fig:results}a with blue and orange lines respectively. To present a transparent analysis, the following theoretical study only considers the contribution to the TINP optical response from $\epsilon_{\parallel}(\omega)$. We also expect a contribution from $\epsilon_{\perp}(\omega)$ (and so should average over all three material axes), but due to limited reliable $\epsilon_{\perp}(\omega)$ data, we present theoretical results derived using only $\epsilon_{\parallel}(\omega)$. Considering the material isotropic in this manner does not affect the conclusions of this work. However for reference, the plot of $\epsilon_{\parallel}(\omega)$ in FIG. \ref{fig:results}a is annotated with vertical lines denoting the $\alpha$ and $\beta$ phonon frequencies for both $\epsilon_{\parallel}(\omega)$ and $\epsilon_{\perp}(\omega)$. For $\epsilon_{\parallel}(\omega)$, the resonance frequencies for the $\alpha$ and $\beta$ phonons are 1.56~THz and 2.85~THz respectively (as given in TABLE \ref{table:bi2te3}), whilst for $\epsilon_{\perp}(\omega)$, the frequencies used are 2.8~THz and 3.5~THZ respectively~\cite{richter1977raman}.
\\

\begin{table}[ht]
\begin{center}
\begin{tabular}{ c c c c } 
 \hline
  & $\quad\omega_{\mathrm{p,j}}/2\pi~\mathrm{(THz)}\quad$ & $\quad\omega_{0,\mathrm{j}}/2\pi~\mathrm{(THz)}\quad$ & $\quad\gamma_{\mathrm{j}}~\mathrm{(THz)}\quad $\\ 
  \hline
 $\alpha\quad$ & 21 & 1.56 & 0.18 \\
 $\beta\quad$ & 4 & 2.85 & 0.2 \\
 $\mathrm{f}\quad$ & 11 & 0 & 0.24 \\
 \hline\\
\end{tabular}
 \end{center}
 \caption{Parameters of $\epsilon_{\parallel}(\omega)$ for Bi$_2$Te$_3$ (calculated from experimental data in \cite{PhysRevB.96.235202}). $\omega_{\mathrm{p,j}}$, $\omega_{0,\mathrm{j}}$ and $\gamma_\mathrm{j}$ denote the amplitude, resonance frequency and harmonic broadening parameters for each mode.}\label{table:bi2te3}
 \end{table}
The energy levels residing in the Dirac cone are evenly spaced, with energies increasing away from the $\Gamma$-point in integer values of $A/R$. $A$ = 2.0~eV\AA $\,$ is a material-dependent quantity, given by the matrix element of momentum calculated for a four band model Hamiltonian of Bi$_2$Te$_3$ and averaged over three axes, calculated by a density functional theory approach and taken from Liu \textit{et al}~\cite{liu2010model} and discussed in more detail in Appendix \ref{app:bands}. 

The absorption cross-section for a spherical particle suspended in a transparent material (in our case, mineral oil for which $n_{\mathrm{oil}}=1.47$ and $\epsilon_{\mathrm{oil}}=2.16$) is calculated from the imaginary component of the particle polarizability $\alpha_0$, such that $\sigma_{\mathrm{abs}} = \frac{k}{\epsilon_0}\mathrm{Im}\{ \alpha_0\}$ and is given by~\cite{siroki2016single} 
\begin{equation}
\sigma_{\mathrm{abs}}(\omega) = 4\pi R^3 n_{\mathrm{oil}}\frac{2\pi}{\lambda}\mathrm{Im}\left[\frac{\epsilon(\omega)+\delta_R(\omega)-\epsilon_{\mathrm{oil}} }{\epsilon(\omega)+\delta_R(\omega)+2\epsilon_{\mathrm{oil}}}\right],
\label{eq:sigma}
\end{equation}
where, for $E_\mathrm{F} = A/R$ (meaning with all states with energy up to and including $A/R$ are occupied), the $\delta_R$ contribution is given by
\begin{equation}\label{eq:delta_AR}
\delta_R(\omega) = \frac{e^2}{3\pi \epsilon_0}\left(\frac{1}{A-\hbar \omega R}+\frac{1}{A+\hbar \omega R}\right).
\end{equation}
The $\delta_R$ term is the contribution to the particle polarizability from the topological delocalised surface states, in which transitions occur between the quantized energy levels of the discretized Dirac cone resulting in a modified surface charge density. The derivation of this term can be found in the supplementary material of reference ~\cite{siroki2016single}. For any Fermi level that resides in the band gap but does not fall within $\pm A/R$, the dominant $\delta_R$ contribution to the absorption cross-section is given by a transition between two energy levels differing by $A/R$. In the absence of surface states (such as by applying a magnetic field term and thus destroying the surface states), $\delta_R = 0$ and we return to the usual solution of a dielectric sphere in a constant electric field. 

For a given Fermi level, $\delta_R$ varies only with $R$ and $\omega$. In FIG. \ref{fig:results}b, we plot the theoretically expected absorption cross-sections of TINPs with $E_\mathrm{F}=A/R$ and radii 20, 25 and 30~nm respectively. Each cross-section has a characteristic shape of three peaks, corresponding to the $\beta$ phonon mode, a localised surface plasmon-polariton (LSPP) mode, and the SToP mode. A point of zero absorption can be seen for each SToP mode, at which point a single electron in a Dirac state shields the bulk from absorbing incoming light. The $R$ and $\omega$ dependence of the absorption cross-section is illustrated in FIG.~\ref{fig:results}d, where the LSPP and $\beta$ modes can clearly be seen, and both the SToP peak and trough of zero absorption can be seen. The $\delta_R$ term in Equation \ref{eq:delta_AR} has no dependence on the dielectric function of either the topological material or the mineral oil, but is affected by the $A$ value. The peak in absorption due to the topological term will experience a shift dependent on which material is being used. For example, the SToP peak for a $R$=20~nm TINP made of Bi$_2$Te$_3$ and Fermi level $E_\mathrm{F}=A/R$ will occur at 2.21~THz, whereas for a Bi$_2$Se$_3$ TINP of the same radius (with A=3.0~eV\AA$\,$  \cite{liu2010model}, as studied in the original work by Siroki \textit{et al}. \cite{siroki2016single}) the peak will occur at 3.1~THz. Sensitivity of $\sigma_{\mathrm{abs}}$ with variation in $A$ is discussed more in Appendix \ref{app:bands}. 

For $-A/R<E_\mathrm{F}<A/R$, the dominant transition is between energy levels directly below and above the Fermi-level, of energy $2A/R$.
\begin{equation}
 \delta_R(\omega) = \frac{e^2}{6\pi\epsilon_0}\left(\frac{1}{2A-\hbar\omega R}+\frac{1}{2A+\hbar\omega R}\right).
 \label{eq:delta_0}
 \end{equation}
FIG.~\ref{fig:results}c shows $\sigma_{abs}/\pi R^2$ for a particle $R=40$~nm, at the two Fermi levels $E_{\mathrm{F}}=A/R$ and $E_{\mathrm{F}}=0$. While the bulk modes remain in the same positions, we see a shift in the SToP mode peaks. The behaviour of the SToP mode and zero absorption trough can be seen in FIG.~\ref{fig:results}e. Although we do not know the precise Fermi level of each TINP in the sample, for particles whose Fermi level resides in the band gap (and thus contribute to the SToP mode peak in the absorption cross-section) it is most likely that transitions are occurring between energy levels separated by $A/R$ rather than $2A/R$.
%
%
\begin{figure}[th!]
\includegraphics[width=\columnwidth]{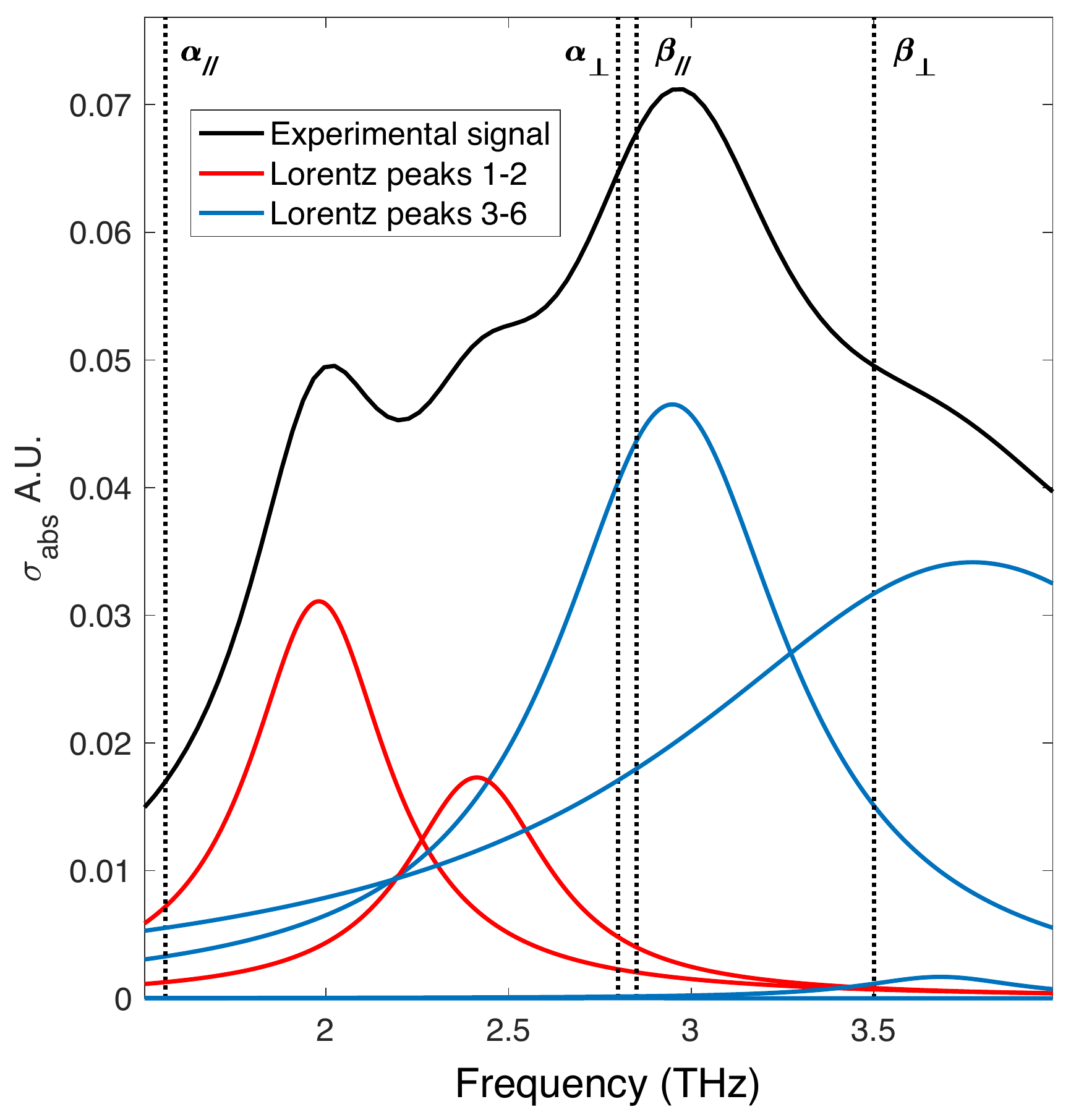}
\caption{\textbf{Experimental results:} Experimental absorption cross-section data with impurity absorption band subtracted, superimposed with a Lorentzian decomposition of the same data, fitted with six peaks. Four prominent peaks are seen in the range 1.5-4 THz, with the first two peaks (in red) presenting potential SToP modes and the last two major peaks presenting probable $\beta$ modes (in blue).} \label{fig:exp_results}
\end{figure}
%
%
\\
\\
In FIG. \ref{fig:exp_results}, we show the experimental absorption cross-section (solid black line) of Bi$_2$Te$_3$ TINPs (of average radius 17.5~nm, standard deviation 1.8~nm) suspended in mineral oil, measured at 300~K. The vertical dotted lines denote the four phonon frequencies ($\alpha_{\parallel}$, $\beta_{\parallel}$, $\alpha_{\perp}$, $\beta{\perp}$) corresponding to the two bulk dielectric functions $\epsilon_{\parallel}$ and $\epsilon_{\perp}$ respectively. By performing a Lorentzian decomposition (using six Lorentz contributions yielding a fit error of 0.16\%, whereas using four, five or seven peaks results in a fit error of 5.18\%, 1.32\% or 31.64\% respectively) it can be seen that two of the Lorentz peaks (in blue) correspond to peaks to the right of each $\beta$ phonon frequency (at 2.95~THz and 3.77~THz respectively, in comparison to $\beta$ phonon frequencies of 2.85~THz and 3.5~THz respectively). 

Crucially, there are two Lorentz peaks (in red) which do not correspond to $\alpha$ or $\beta$ modes for any material axis. It cannot be entirely excluded that one of the peaks is the LSPP mode from the $\epsilon_{\perp}$ contribution to the absorption, due to low confidence in the available data. However the origin of at least one of the peaks cannot be attributed to bulk properties.  $\alpha$ modes are typically over-damped and so are dark modes and are not seen in the absorption cross-section. This can be seen easily from the dielectric function for Bi$_2$Te$_3$ depicted in FIG. \ref{fig:results}a, where the very large value of Im$[\epsilon (\omega)]$ at the $\alpha$ phonon frequency will result in a very small contribution to $\sigma_{\mathrm{abs}}$ (given in equation \ref{eq:sigma}) at this frequency, plotted in FIGs~\ref{fig:results}b-e .

No bulk modes are expected in the frequency range between the $\alpha_{\parallel}$ and $\alpha_{\perp}$ frequencies, and so we conclude that at least one of the unexpected peaks observed in the experimental are contributions from the topological surface states. The presence of two unexpected peaks is most likely due to separate resonances from incoming light aligning randomly with the axes of the Bi$_2$Te$_3$ crystal structure so there will be contributions from both $\epsilon_{\parallel}$ and $\epsilon_{\perp}$, although as previously explained, one of the peaks could be the LSPP mode from $\epsilon_\perp$. The modes seen in the experiment are much less visible than those found in theory, in part due to temperature smearing as the experiment operates at room temperature ($\sim$ 6~THz), where the probability of finding a level empty in our system it is low. This could be remedied by studying a TI with a bigger band gap and/or smaller particles. Due to this and other experimental considerations (such as spread of $R$ values for the nanoparticles in the sample, non-spherical particles - which is studied in more detail in Appendix \ref{app:cube}) all of which will cause minor shifts or spreading of peaks in the absorption cross-section, the simple theory model in this paper is not expected to capture the quantitative results of the experiments, but qualitative agreement appears to be good. It should also be noted that in our theory approach we consider a single electron problem, and although this should be a good approximation due to the high level of delocalization of the electrons on the TINP surface, some deviation can be expected.
\\
\\
We conclude from this experimental comparison with the theory model that we have successfully observed the SToP mode in Bi$_2$Te$_3$ nanoparticles, which was theoretically predicted in~\cite{siroki2016single}. SToP modes represent the realization of topological quantum dots that could have a disruptive role in quantum optics and as THz sources.

\section*{Acknowledgements}
M.S.S. would like to acknowledge the President’s PhD Scholarship programme at Imperial College London for financial support. C.M. would like to  acknowledge the EPSRC award EP/M022250/1, and the award of a Royal Society University Research Fellowship by the UK Royal Society. M.S.R. is supported through a studentship in the Centre for Doctoral Training on Theory and Simulation of Materials at Imperial College London funded by EPSRC Grant No. EP/L015579/1.
V.G. acknowledges the Spanish Ministerio de Economia y Competitividad for financial support through the grant NANOTOPO (FIS2017-91413-EXP), and also Consejo Superior de Investigaciones Científicas (INTRAMURALES 201750I039).
\bibliographystyle{unsrt}
\bibliography{references}

\appendix
\begin{figure}
\includegraphics[width=\columnwidth]{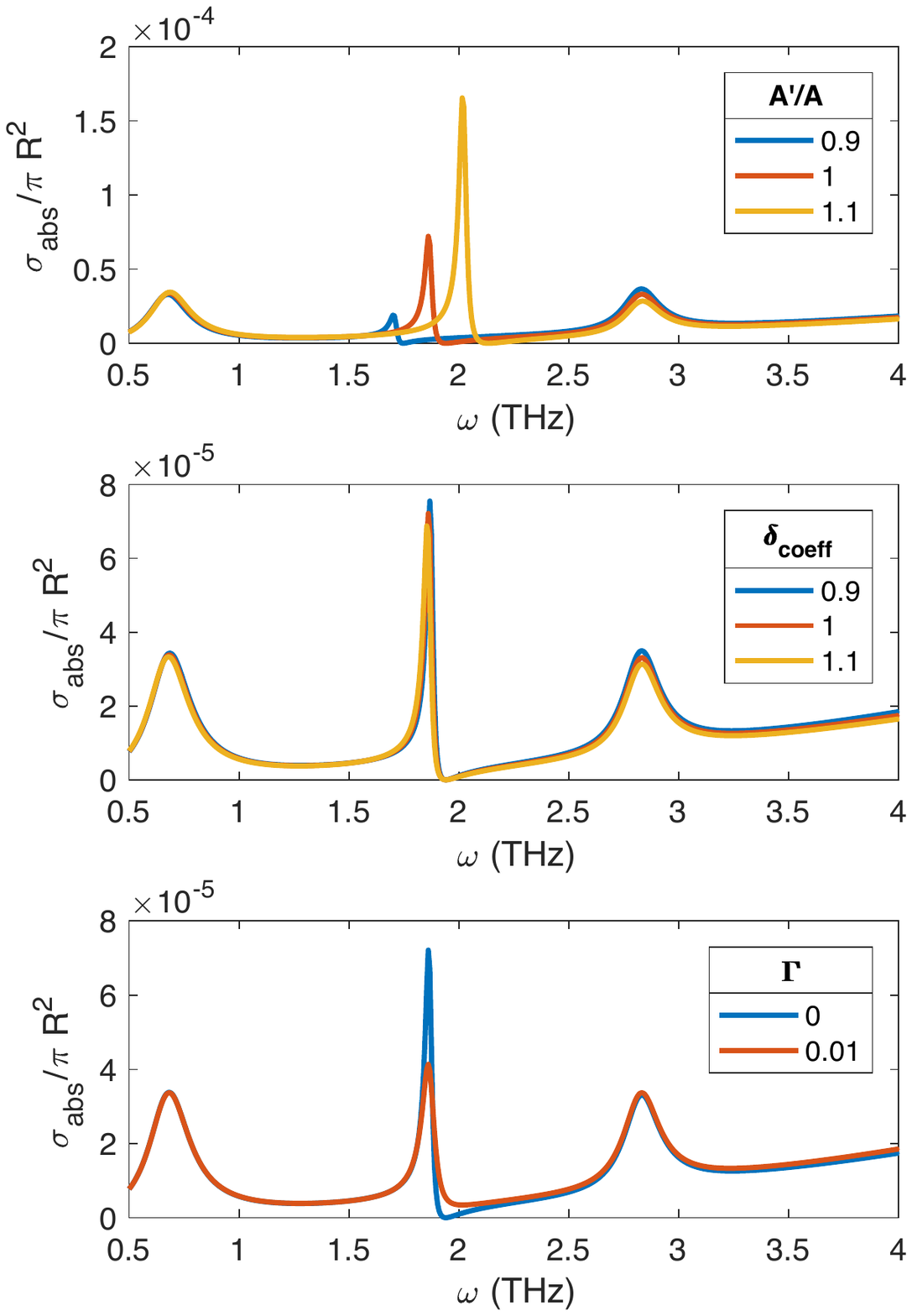}
\caption{\textbf{Modifying theory parameters:} \textbf{(a)} Varying $A$ by 10\%, which shifts the position of the SToP mode peak and modifies the height of the peak. \textbf{(b)} Varying the magnitude of $\delta (\omega)$ by 10 \%, which has no effect on the position of the SToP mode peak and very little effect on the height of the peak. \textbf{(c)} Introducing a small complex component to the denominator of $\delta(\omega)$, equivalent to introducing a finite life time to the excited surface states. Introduction of this finite lifetime reduces the height of the SToP mode peak.  \label{fig:modify}}
\end{figure}
%
\begin{figure}
\includegraphics[width=\columnwidth]{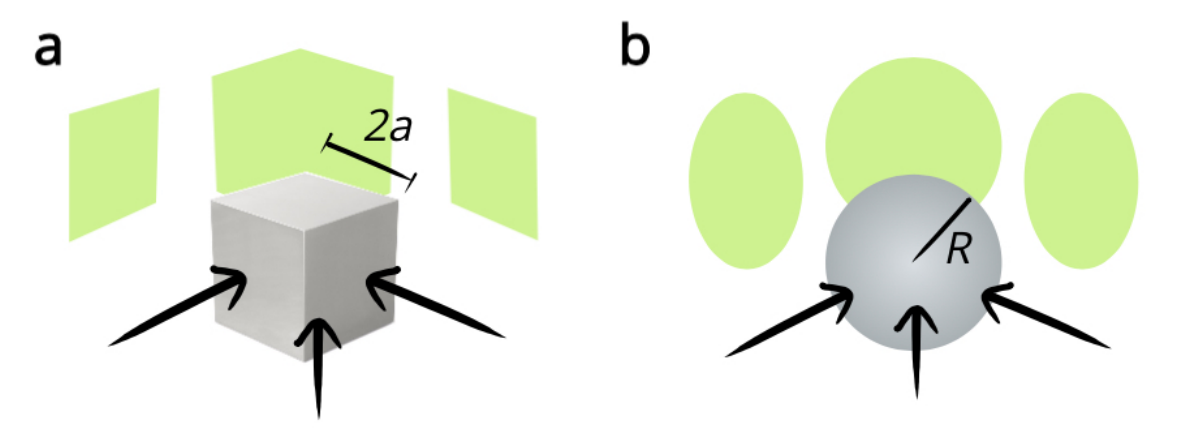}
\caption{\textbf{Average projections of nanoparticles:} \textbf{(a)} Projection of a cube of side length $L=2a$ for differing directions. Minimal projection will be the surface of one side, $4a^2$ and average projection is $6a^2$. \textbf{(b)} Projection of a sphere of radius $R$ has both minimal and average projection of $\pi R^2$. For particles of the same volume, the cube will have a larger average projection. \label{fig:projections}}
\end{figure}
%
\section{Experimental method}\label{app:exp}
In a standard synthesis, 114 mg of bismuth acetate (Bi(CH$_3$COO)$_3$, 99\% Aldrich) and 3.5~mL of 1-dodecanethiol (DDT, 98\% Aldrich) were mixed in a three-neck flask and heated to 45$^{\circ}$C under vacuum and kept at this temperature until a transparent pale-yellow solution is formed. Then the flask was flashed with nitrogen and heated to 60$^{\circ}$C and 6.5~mL of oleylamine (OlAm, 70\%, Aldrich) was quickly added. After 24 hours the as-prepared bismuth nanoparticles were used without any further purification. For tellurisation 0.45~mL of 1M trioctylphosphine telluride (TOP:Te) was injected at 60$^{\circ}$C into the solution containing bismuth nanoparticles. The reaction mixture was kept at this temperature for 48 hours until complete tellurisation and then annealed at 110$^{\circ}$C for 8 hours in order to restore crystallinity. Thus produced Bi$_2$Te$_3$ nanoparticles were washed three times with ethanol and then redispersed in chloroform. For THz measurements solvent was exchanged to mineral oil.
%
\section{Four band model Hamiltonian}\label{app:bands}
The important topological physics occurs near the $\Gamma$-point, allowing us to use \textbf{k}$\cdot$\textbf{p} perturbation theory to expand about the $\Gamma$-point and write down a low-energy, 4-band effective Hamiltonian,
\begin{equation}
    H(\mathbf{k}) = E_0 (\mathbf{k})\mathbb{1}_4 + \begin{pmatrix}\mathcal{M}(\mathbf{k}) & A_1 k_z & 0 & A_2 k_- \\ A_1 k_z & -\mathcal{M}(\mathbf{k}) & A_2 k_- & 0 \\ 0 & A_2 k_+ & \mathcal{M}(\mathbf{k}) & -A_1 k_z \\ A_2 k_+ & 0 & -A_1 k_z & -\mathcal{M}(\mathbf{k})\end{pmatrix}
\end{equation}
where $k_{\pm} = k_x \pm i k_y$, $E_0 (\mathbf{k}) = C + D_1 k_z^2 + D_2 k_{\perp}^2$ and $\mathcal{M}(\textbf{k}) = M-B_1k_z^2 -B_2 k_{\perp}^2$. The parameters ($A_1$, $A_2$, $B_1$, $B_2$, $C$, $D_1$, $D_2$, $M$) in this effective model can be determined by fitting the energy spectrum of the Hamiltonian to that of \textit{ab initio} calculations~\cite{zhang2009topological}, with the results for Bi$_2$(Se$_x$Te$_{1-x}$)$_3$ presented by Liu \textit{et al}~\cite{liu2010model}. In the theory model used in this work, the crystal structure is taken to be isotropic, such that $A = \frac{1}{3}(A_1+2A_2)$. FIG. \ref{fig:modify}a shows how the theoretically calculated absorption cross-section varies with uncertainty in $A$. Shifting $A\rightarrow A'$ by 10\% shifts the position of the SToP peak by $< 10\%$. FIG.~\ref{fig:modify}b illustrates that modifying the magnitude of the topological contribution  to the particle polarizability $\delta_{R}(\omega)$ (by multiplying by a constant coefficient $\delta_{\mathrm{coeff}}$),) has little effect on the position of the SToP peak. FIG.~\ref{fig:modify}c shows that when a finite lifetime of excited surface states is considered (by detracting a small, complex component $i\Gamma A$ from the denominator of all $\delta_R(\omega)$ terms), the height of the SToP mode peak becomes smaller (for a particle of $R=25$ nm and $\Gamma = 0.01$, the height of the SToP mode peak reduces by $\sim 45\%$) and the position remains the same.  
\\
\section{Modification to absorption cross-section for non-spherical particles}\label{app:cube}

The analytic formula for the absorption cross-section of a cube can be compared to that of a sphere of  the same volume~\cite{massa2013analytical}, but we found that the position of the SToP mode was very resilient to this perturbation, which is consistent with the behaviour we would expect from the delocalised surface states. However this calculation relies on light hitting the surface of the particle along a single axis of the material. When an average orientation of the particle is considered, the results for sphere and cube are expected to differ slightly.

The absorption cross-section of a particle of arbitrary orientation w.r.t incoming light depends on its average projection (which in generality for a concave shape is equal to $\frac{1}{4}$ of its surface area). The average projection of a sphere is smaller than the average projection of a cube of the same volume (i.e. for $V_{\mathrm{sphere}} = \frac{4}{3}\pi R^3 = V_{\mathrm{cube}}=8a^3$, $P_\mathrm{cube}\approx 1.24 P_{\mathrm{sphere}}$), see FIG.~\ref{fig:projections}. The experimentally synthesized Bi$_2$Te$_3$ nanoparticles in this work are slightly rhombohedral, and so may be more accurately described by cubes. By comparing the experimental results of a cube to the theoretical model for a sphere used throughout this work, it should be noted that a cube of side length $L=2a$ will give the same average projection area of a sphere of radius $R\approx 1.36a$. This in part explains the discrepancy between the experimentally measured size distribution of Bi$_2$Te$_3$ particles and the position of the observed SToP mode peaks.

\end{document}